\documentclass[%
 reprint,
 amsmath,amssymb,
 aps,
]{revtex4-1}

\usepackage{dcolumn}
\usepackage{bm}
\usepackage{graphicx,times}             
\usepackage{natbib}
\usepackage{amssymb,amsmath}
\usepackage{longtable}
\usepackage{rotating}
\usepackage{multirow}
\usepackage{diagbox}
\usepackage{hyperref}
\usepackage{xcolor}

\begin{document}

\preprint{APS/123-QED}

\title{Calibrating the effective magnitudes of type Ia supernovae with a model-independent method}

\author{Jian Hu} \email{dg1626002@smail.nju.edu.cn, corresponding author}
\altaffiliation{Institute of Astronomy and Information, Dali University, Dali 671003, China}

\author{Jian-Ping Hu} \email{hjp2022@nju.edu.cn, corresponding author}
\altaffiliation{School of Astronomy and Space Science, Nanjing University, Nanjing 210093, China}
\author{Zhongmu Li}
\altaffiliation{Institute of Astronomy and Information, Dali University, Dali 671003, China}
\author{Wenchang Zhao}
\altaffiliation{Institute of Astronomy and Information, Dali University, Dali 671003, China}
\author{Jing Chen}
\altaffiliation{Institute of Astronomy and Information, Dali University, Dali 671003, China}

\date{\today}

\begin{abstract}
This research explores the correlation between the absolute magnitude and the redshift of Type Ia supernovae (SNe Ia) with a model-independent approach. The Pantheon sample of SNe Ia and strong gravitational lensing systems (SGLS) are used. With the cosmic distance-duality relation (CDDR), the  evolution parameter of the magnitude, the light curve parameters of SNe Ia, and the parameters of the SGLS geometric model are constrained simultaneously. Considering the consistency of the redshifts, we selected a subsample of SNe Ia in which the redshift of each SNe Ia is close to the corresponding redshift of the SGLS sample. Two parametric models are used to describe this evolution, which can be written as $\delta_M=\varepsilon z$ and $\delta_M=\varepsilon\log(1+z)$, respectively. Our analysis reveals that $\varepsilon=-0.036^{+0.357}_{-0.339}$ in the first parametric model and $\varepsilon=-0.014^{+0.588}_{-0.630}$ in the second model, indicating that no significant evolution ($\varepsilon=0$) is supported at the 1$\sigma$ confidence level in this study. These results represent a significant advancement in our understanding of the intrinsic properties of SNe Ia and provide important constraints for future SNe Ia study.


\end{abstract}

\maketitle


\section{Introduction}           
\label{sect:intro}

In 1998, the accelerating expansion of the universe was discovered by measuring the relation between redshift and the distance of SNe Ia. As a powerful tool in cosmology, SNe Ia has a common origin and approximately equal luminosity, making it a potential standard candle. To use SNe Ia as a standard candle, standardization is required, through empirical procedures based on their light-curve shape and color \citep{phi93,rie96,per97,guy07,jha07}. Because the light curves  of the SNe Ia are different, they must be normalized(e.g. the Phillips relationship, \cite{phi93,phi99} ) before these SNe Ia can provide reliable luminosity distances to studying the cosmology. For example, the equation of state of the dark energy\citep{hute99, hann02,wan07,wan14}, the cosmology curve\citep{bet14,ras15,col19}, the cosmology opacity\citep{avgo09,lizh13,hola14,hyw17}, the CDDR\citep{baku04, hol10, njj11, hlr11, cali11, mzz12, huji18, hu23}, and so on. However, these studies might neglect the evolution of the absolute magnitude of SNe Ia. Despite most of SNe Ia coming from the same mechanism, the metallicity, mass, and other features of the SNe Ia progenitors may be different, which would affect the SNe Ia magnitudes. If we neglect the effect of progenitor properties on SNe Ia magnitudes in cosmology tests, it might cause an additional systematic error.

More than a decade ago, some researchers began to study this problem. \citet{wrig02} found that the SNe Ia magnitude evolution with an exponential function of cosmic time in the Einstein de Sitter cosmology model may mimic dark energy. \cite{nor08} use a simple linear form ($\delta_M=kz$, where $k$ is the slope) to estimate the bias of the luminosity distances with the Gold sample SNe Ia which is presented in \citet{rie07}, and they found that this effect may cause significant bias only if the slope $k$ is $\sim |0.1|$.

More recently, several more researchers have examined this question. \citet{kim19} use their constructed SNe sample from YONSEI (Yonsei Nearby Supernova Evolution Investigation) to explore the relation between the magnitude of SNe Ia and redshift. They found that SNe Ia low redshifts are about 0.07$\sim$ 0.08 magnitude fainter than those in high redshift by eliminating the effect of the galactic environment. \citet{kan20} found a significant correlation between SNe standardization luminosity and stellar population age at a 99.5\% confidence level.

In terms of testing methods, \citet{lind09} explored the effect of SNe Ia magnitude evolution in fitting cosmological parameters. They assume that the peak magnitude of SNe Ia evolved with cosmic age in a linear function and found that the magnitude evolution of SNe Ia in high-redshift appears to be some percent brighter than would be expected in a standard cosmology model. In similar work published by \citet{tut17}, they found that the conclusion of cosmic accelerated expansion is not clear if the independence between SNe Ia intrinsic luminosity and its redshift is not imposed. So, parametric models independent of a cosmological theory method to test the evolution of the magnitude of SNe Ia is crucial for this topic. \citet{evsl16} developed a model-independent method that uses the ratio of two angular baryon acoustic oscillation (BAO) scales at redshifts 0.32 and 2.34 from BOSS BAO to calibrate the SNe Ia magnitude. The result shows that a statistically insignificant downward shift with $M_B(2.34)-M_B(0.32)= -0.08\pm 0.15$ for JLA SNe Ia at low z and Hubble Space Telescope SNe Ia at $z > 1.7$, and a shift of $-0.24 \pm 0.13$ for BOSS data with the best fit Planck $\Lambda$CDM BAO expectations. Their results would have been more reliable if they had simultaneously fitted the parameters of the supernova's light curve and magnitude evolution , instead of using someone else's light curve to calculate this parameter, which did not consider the evolution of the absolute magnitude. \citet{zha19} use the data of gravitational-wave sources GW170817 electromagnetic (EM) counterparts to calibrate the absolute magnitude of SNe Ia. It is a cosmological model-independent method, but only a few data sets are available for his work. \citet{wen20} uses the $D_L$ from gravitational wave signals and the ratio of $D_A$ from the SGLS geometric model with a cosmology model-independent method, which is proposed by \citet{lia16}. On the subject of constraining the absolute magnitude of SNe Ia, although their method is also model-independent, it does not parameterize the $M_B$ and only precisely limits the value of it.

In this article, we explore whether this effect is significant using an approach that is independent of cosmological models, which may avoid the above problem. The method is to join the SNe Ia and strong gravitational lensing systems and use the strong lens model to replace the cosmology model to fit the SNe Ia light curve parameters. The strong lensing sample and a selected SNe Ia sample are taken. To avoid introducing additional parameters, the redshift of the SNe sample is close enough to the strong lensing sample, which can be regarded as the same luminosity distance.

This paper is organized as follows. In the next section, we show the data and the method of selecting the data. In section 3, the fitting method and the numerical results are shown. Section 4 gives the conclusions.


\section{Data}
\label{sect:data}

In this section, we will introduce the data used in this work. The data include the latest SNe Ia sample and the SGLS samples. We also present the technique we employ for matching data to acquire two different distance measures at identical redshift.

\subsection{The sample of SNe Ia}\label{II.A.}
In this study, we employ data from the pantheon sample and the fitting method from \citet{sco18}. Pan-STARRS1 (PS1) Medium Deep Survey, Sloan Digital Sky Survey (SDSS), SNLS, and several low-z and Hubble Space Telescope samples compose the sample. This collection contains 1048 SNe Ia with redshifts between 0.01 and 2.3.
In their work, they adopt a common distance modulus form $\mu=5\log_{10}(D_L/\rm M pc)+25$ may be expressed as follows \citep{tri98}:

\begin{equation}\label{211}
	\mu=m_B^ \star-M_B+\alpha \times x_1-\beta \times c+\Delta_M,
\end{equation}
where $m_B^ \star$ refers to the measured peak magnitude in the B band rest frame, $x_1$ denotes the light curve stretch parameter, and $c$ denotes the light curve color parameter, which are usually different for different SNe Ia. $M_B$ is the B-band fitting absolute magnitude of a fiducial SNe Ia with $x_1=0$ and $c=0$. $\alpha$ and $\beta$ are nuisance distance estimate parameters. $\Delta_M$ is a distance correction depending on the mass of the SNe's host galaxy. In the research of \citet{sco18}, it was parameterized by
\begin{equation}\label{212}
	\Delta_{M}=\gamma \times\left[1+e^{\left(-\left(m-m_{\text {step }}\right) / \tau\right)}\right]^{-1}],
\end{equation}
$m$ is the observed value, indicating the $\log_{10}$ mass of the host galaxy of the SNe Ia, $\gamma$, $m_{\text {step }}$ and $\tau$ are fitting parameters.
To explore the evolution of SNe Ia absolute magnitude, a tiny function $\delta_M$ about redshift is introduced:
\begin{equation}\label{213}
	M_B=M_{B0}+\delta_M.
\end{equation}
Here $M_B$ is divided into two parts, where $M_{B0}$ is the absolute magnitude of SNe Ia at $z=0$, and $\delta_M$ is part of the absolute magnitude that evolves with redshift. If $\delta_M=0$, it means that $M_B$ does not evolve with the redshift, so $M_{B0}$ is $M_{B}$.

In this work, we parameterized the $\delta _M(z)$ with one index parameter. We are parameterizing $\delta _M(z)$ whose value stays zero when $M_B$ is not evolution. Any significant evolution from absolute magnitude might indicate the emergence of a new SNe physics mechanism. Since abundant successful cosmic tests use no evolution assumption of $M_B$, we expect $\delta _M(z)$ to stay close to zero. To simulate any deviation from zero, we parameterize $\delta _M(z)$ using 2 parameter representations to describe the possible redshift dependence of the evolution of absolute magnitude:
\begin{equation}\label{214}
	\delta_M=\varepsilon z,
\end{equation}
and 
\begin{equation}\label{215}
	\delta_M=\varepsilon \log(1+z).
\end{equation}
The two parameterizations fit differently for low and high redshift samples, respectively, with the $\varepsilon$ parameter in equation (\ref{214}) being more sensitive to changes in redshift. Whereas equation (\ref{215}) is more suitable for fitting data whose samples contain very high redshifts. Both parameterizations are independent of the cosmological model. If $\varepsilon=0$ means that the absolute magnitude of type SNe Ia does not evolve with redshift, $\varepsilon<0$, SNe Ia at higher redshifts  are brighter, and $\varepsilon>0$, SNe Ia at higher redshifts are fainter, we would expect the result to be $\varepsilon<0$ or close to zero.

\subsection{Gravitational Lensing}
Our SGLS sample is made up of 205 SGLS from \citet{ama20}, which includes the survey projects SLACS, CASTLES survey, BELLS, and LSD, among others. The theory of general relativity predicts several astronomical occurrences. Gravitational lensing is one of the few most successful predictions among them. Gravitational lensing is the deflection of light radiated by an astronomical body when it passes close to a massive celestial body as a result of gravity. This effect is powerful enough to generate various images, arcs, and even Einstein rings, as the foreground lens is typically a galaxy or galaxy cluster. The SGLS effect is gaining importance as a potent tool for investigating both the background cosmology and the nature of galaxies functioning as lenses. Simultaneously measuring the Einstein ring radius and the star core velocity dispersion $\sigma_{0}$ can yield the angular diameter distance ratio(\textbf{$R^{A}$}):
\begin{equation}\label{221}
	R^{A}(z_{l},z_{s})=\frac{D^A_{ls}}{D^A_{s}},
\end{equation}
$z_{l}$ and $z_{s}$ in equation(\ref{221}) denote the redshift of the lens and source, respectively. Equation(\ref{221}) is the define of the angular diameter distance ratio, where $D^A_{ls}$ and $D^A_{s}$ are, respectively, the angular diameter distances between lens and source and between observer and source. Considering a Singular Isothermal Sphere (SIS) lens model, the distance ratio $R^A$ is associated with observable values as follows \citep{bie10}:
\begin{equation}\label{222}
	R^{A}_{SGLS}(z_{l},z_{s})=\frac{c^{2}{\theta }_{E}}{4\pi\sigma^2_{SIS}}.
\end{equation}
In the above equation, $c$ represents the speed of light, $\theta _E$ represents the Einstein radius,  $\sigma_{SIS}$ represents the velocity dispersion generated by the lens's mass distribution, and the subscript ``SGLS" indicates that the value of expression is provided by the data of SGLS. In general, we must remark that $\sigma_{SIS}$ is not precisely equal to the measured star velocity dispersion $\sigma_{0}$ \citep{whi96}. To account for this disparity, the phenomenological free parameter $f_e$ is created and defined as $\sigma_{SIS} = f_e\sigma_{0}$, where $0.8^{0.5}<f_e< 1.2^{0.5}$ \citep{ofe03}.

For the more general case of deviation from the standard ``SIS" model, since the slope of the density curve of a single galaxy deviates from $2$ (i.e. the standard SIS model), assuming that its density distribution is a spherically symmetric power-law distribution of the type $\rho\sim r^{-\gamma}$, the Einstein radius can be written as \citep{cao15}
\begin{equation}\label{2208}
	\theta_E=4\pi {D_{A_{ls}}\over D_{A_{s}}}\frac{\sigma_{ap}^2}{c^2}\bigg({\theta_E\over \theta_{ap}}\bigg)^{2-\gamma}f(\gamma),
\end{equation} 
where $\theta_{ap}$ denotes the stellar velocity dispersion within an aperture of size ap and $f(\gamma)$ is written as
\begin{equation}\label{2209}
	f(\gamma)=-\frac{(5-2\gamma)(1-\gamma)}{\sqrt{\pi}(3-\gamma)}\frac{\Gamma(\gamma-1)}{\Gamma(\gamma-3/2)}\bigg[\frac{\Gamma(\gamma/2-1/2)}{\Gamma(\gamma/2)}\bigg]^2\,.
\end{equation}
According to researchers's previous work on the $\gamma$ index of the SGLS, the deviation of this index from $2$ is usually small. For example, \cite{hola17} used the (Union2.1) SNe Ia sample and the then latest 167 GRBs as luminosity distance metrics to calibrate the SGLS $\gamma$ indices, which are in the form of indices containing two parameters $\gamma_0$ and $\gamma_1$, and they divided the samples in various ways, resulting in all subsamples and the total sample supporting the point ($\gamma_0$, $\gamma_1$) = (1, 0) at the 1$\sigma$ confidence region. \cite{cui17} used the SNe Ia sample of JLA, the CMB sample, the BAO sample, and the H(z) sample to constrain the $\gamma$ index of SGLS and obtained similar conclusions. So we just set $\gamma=2$ here. The star velocity dispersion $\sigma_{0}$ can be expressed as
$\sigma_{ap}(\theta_{eff}/(2\theta_{ap}))^{-0.04}$ \citep{jor95a,jor95b}, which has already been dealt with in the work of \citet{ama20}. Therefore equation(\ref{2208}) becomes equation(\ref{222}).

In a model of flat cosmology, the comoving distance(($D^c(z)=(1+z)D^A$)) between the lens and the source is expressed as $D^c_{ls}=D^c_s-D^c_l$. The ratio of the comoving angular diameter distance may thus be rewritten as follows:
\begin{equation}\label{223}
	R^{A}(z_{l},z_{s})=1-\frac{(1+z_{l})D^A_{l}}{(1+z_{s})D^A_{s}}=1-\frac{D^c_l}{D^c_s}=1-\frac{d^c_l}{d^c_s},
\end{equation}
where $d^c_l$ and $d^c_s$ are the dimensionless distances $d =\frac{H_0}{c}D^c$.

\subsection{Pair off data of SNe Ia and SGLS}\label{II.C.}

To fit these parameters, the redshift of SNe Ia must be near the source or the SGLS lens. In other words, an SGLS requires two SNe Ia to couple the source and lens independently. To make full use of the data sample, one must couple the subsamples well. We couple subsamples using a distance-deviation consistency technique, which outperforms the usual method that sets a fixed $\Delta z=$constant. For example, $\Delta z=0.005$ \citep{hol10,hga12, li11, njj11},  $\Delta z=0.006$ \citep{ghj12}, and $\Delta z=0.003$ \citep{lia19}.

\begin{figure}
	
	\centering
	\includegraphics[scale=0.6,angle=0]{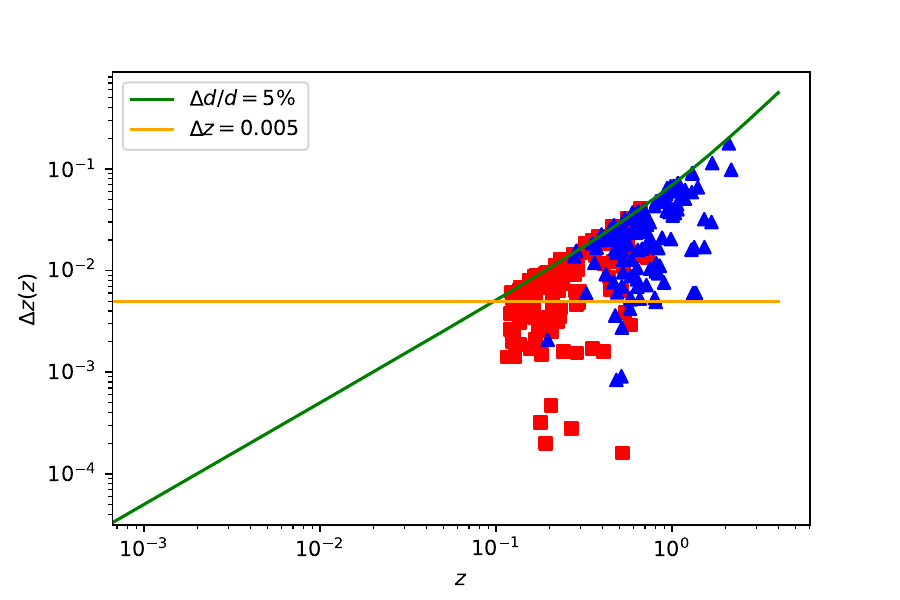}\\
	\caption{The green solid line represents the relation between the $\Delta z$ and $z$ in $\Lambda$CDM with $\Delta d/d=5\%$. The orange solid line represents $\Delta z=0.005$. The red squares indicate the absolute difference between the lens redshift of a SGLS and the redshift of a pair of SNe Ia. The blue triangles indicate the absolute difference between the redshift of the source of a SGLS and that of a paired SNe Ia.}
	\label{fig01}
\end{figure}

In this study, the redshift difference between subsample pairs is not set and decreases with redshifts to ensure that the distance deviation of the sources remains constant.
Following the approach of \citep{zhou21}, the relationship between $z$ and dimensionless distance for a cosmological model may be expressed as follows:
\begin{equation}\label{231}
	\Delta{d_c^{model}(z)}=d_c^{model}(z+\Delta z^{model})-d_c^{model}(z),
\end{equation}
The exact expression of equation(\ref{231}) will vary with the model. In cases when the flat $\Lambda$CDM model with $\Omega_m=0.31$\citep{planck} is used. Setting $\Delta d_c/d_c$ equals 5\% and combining equation(\ref{231}) yields $\Delta{z}(z)$:
\begin{equation}\label{15}
\begin{split}
\int_{0}^{z+\Delta z}\frac{dz^{\prime}}{\sqrt{\Omega_m(1+z^{\prime})^3+1-\Omega_m}}=\\
(1+\Delta d_c/d_c)\int_{0}^{z}\frac{dz^{\prime}}{\sqrt{\Omega_m(1+z^{\prime})^3+1-\Omega_m}},
\end{split}
\end{equation}
which may be used to rule out excessively high selection uncertainties.
By simplifying equation(\ref{231}), we can obtain the ordinary differential equation for $\Delta z$ with respect to $z$, which can be written by:
\begin{equation}\label{024}
	\frac{d\Delta z(z)}{dz}=1.05 \frac{\sqrt{0.31(1+z+\Delta z(z))^3+0.69}}{\sqrt{0.31(1+z)^3+0.69}}-1,
\end{equation}
with a initial value $\Delta z(0)=0$. The numerical solution of equation (\ref{024}) is plotted on the solid green line in Figure (\ref{fig01}). For comparison, we present the  $\Delta z=0.005$ line, represented by the solid orange line in Figure (\ref{fig01}).
As the selection method that meets the requirement is not unique, we determine the sum of the redshift deviations of these paired samples to choose the best paired sample, which must possess a minimum value of $\sum\limits_{i = 1}^{2n} {(\Delta z_i^k)^2}$, where $n$ denotes the logarithmic number of pairings and $k$ represents the $k$ alternatives. Please note that the pairing data utilized here relies on a distinct cosmological model. However, this model only offers a maximum fixed deviation of $d$, which is not taken into consideration in the subsequent calculations and is exclusively implemented to eliminate significant $\Delta d$.

By using this route, not only might more observations at high redshifts be kept, but more conforming data sets could be matched as well, for example, those data in the upper right corner of Figure (\ref{fig01}). If we use our technique with a fixed $\Delta d/d = 5\%$, the number of matched data sets can increase to 68, which is an increase in the utilization of the original data of 13.3\% compared to the work of \citet{lia16}, which utilized a fixed $z=0.005$, which matched 60 data sets. We compiled a total of 120 data points, the maximum redshift of which was $z=2.16$ with fixed $\Delta d/d = 5\%$.

Setting $\Delta d/d = 5\%$ has the following considerations: first, the lowest redshift in our selected SNe Ia is roughly at $z = 0.1$, and the selection using a similar method to \citet{lia16} is also at $z = 0.1$, where the $\Delta d/d$ calculated using the $\Lambda$CDM model is about 5\%. So for the distance-deviation consistency, we take an approach where the upper limit of the selection uncertainty is set to 5\% at all redshifts in the sample, otherwise, at all redshifts, our selecting uncertainty is necessarily larger than the selection uncertainty caused by the previous method. If we fail to match at an uncertainty range of less than or equal to 5\%, we are discarding that SGLS data. We need to increase this selection error up to 5\% so that we can match as much data as possible; for this reason, we will choose a $\Delta d/d$ deviation of 5\% as a compromise.

We made the following considerations for accuracy: The previous method of number selection at high redshifts, such as at $z = 1$, resulted in $\Delta d/d<<1\%$ being significantly smaller than 1\%, indicating a small selection uncertainty at high redshifts but a relatively large selection uncertainty at low redshifts. The uncertainty induced by our technique is constant and corresponds to the uncertainty of the prior method at low redshifts. Our selection uncertainty is significantly higher than that of previous research at high redshifts, but this does not necessarily mean that our results are less precise than those of prior research. Their approach failed to account for this component of the error in a model-independent method, as the selection error is $z$ and the implicated variable or intermediate variable is $d$. Linking $d$ to $z$ requires the use of a specific cosmological model. Therefore, their approach explicitly disregards this choosing uncertainty, which is contentious. Their method makes it simpler to match more data at low redshifts, despite this component having a rather substantial selection uncertainty, while the high redshift is difficult to match despite the minimal selection uncertainty. Their method neglects the large deviation and discards the data at high redshifts, while our method resolves this problem.

\section{Simulatons and results}
\subsection{Simulatons}
In this study, we employed a model-independent technique to investigate the evolutive parameter of SNe Ia magnitude. For this technique, the geometric model of strong lensing replaces the cosmological model in fitting the SNe Ia light curve in the 2 to its parameters. Then, the light-curve parameters are connected with the strong lensing model's parameters. To investigate the potential evolution of the magnitude of SNe Ia, we use two parameterized forms equation(\ref{214}) and equation(\ref{215}). Because the redshifts of some SGLS and SNe Ia are quite large, \textbf{a} model for redshift nonlinearity will also be adopted.

The two diameters in equation(\ref{223}) should be computed using a model-independent approach. To get these distances, the CDDR and luminosity distance of SNe Ia must be taken into account. The CDDR, also known as the Etherington relation, is a crucial concept in observational cosmology. The following equation connects the luminosity distance ($D_L$) to the angular diameter distance ($D_A$)
\begin{equation}\label{311}
	D_L=D_A(1+z)^2.
\end{equation}
It is valid for all Riemannian geometry-based cosmological theories. The theoretical foundations of this relationship are the conservatism of the photon count and the photons' motion along the null geodesics in Riemannian space-time \citep{ell07}. This relationship plays a crucial role in modern cosmology, particularly in galaxy observations\citep{cmsl07, man10}, gravitational lensing \citep{ell07}, and cosmic microwave background (CMB) radiation observations\citep{kom11}.

To gain the value of parameters, we must construct  the $\chi^2$ analysis to the assumed parameterizations' parameters. When equation(\ref{223}) and equation(\ref{311}) are combined, the angular diameter distance ratio may alternatively be expressed as 
\begin{equation}\label{312}
	R^{A}_{SNe}(z_{l},z_{s})=1-\frac{(1+z_{s})D_{Ll}}{(1+z_{l})D_{Ls}},
\end{equation}
where $D_{Ll}$ and $D_{Ls}$  are the  luminosity distance of lens and source of SGLS, respectively, and the subscript ``SNe" indicates that the value of expression is provided by the data of SNe Ia. They can unitedly express by:
\begin{equation}\label{313}
	D_{L}=10^{(0.2(m_B^ \star-M_B+\alpha \times X_1-\beta \times C+\Delta_M)-5)}.
\end{equation}

Combining equation(\ref{211}) and equation(\ref{312}) with $\mu=5\log10(D_L/\rm M pc)+25$ yields the form of $\chi^2$:
\begin{equation}\label{314}
	\chi^2=\sum\frac{(R^{A}_{SNe}(z_{l},z_{s})-R^{A}_{SGLS}(z_{l},z_{s}))^2}{\sigma^2_{R^{A}_{SNe}}+\sigma^2_{R^{A}_{SNe,sel}}+\sigma^2_{R^{A}_{SGLS}}},
\end{equation}
$\sigma_{R^{A}_{SNe}}$ is the uncertainty of $R^{A}_{SNe}(z_{l},z_{s})$ and may be computed as follows: 
\begin{equation}\label{315}
	\sigma_{R^{A}_{SNe}}=(1-R^{A}_{SNe}(z_{l},z_{s}))\sqrt{\sigma^2_{D_{Ll}}/D^2_{Ll}+\sigma^2_{D_{Ls}}/D^2_{Ls}},
\end{equation}

$\sigma_{D_{Ll}}$ and $\sigma_{D_{Ls}}$ are the respective uncertainty of ${D_{Ll}}$ and ${D_{Ls}}$. According to equation(\ref{313}), it can also use a united form to express by:
\begin{equation}\label{316}
	\sigma_{D_{L}}=\frac{\ln(10)}{5}D_{L}\Delta \mu.
\end{equation}
$\sigma_{R^{A}_{SNe,sel}}$ is the selection uncertainty of $R^{A}_{SNe}(z_{l},z_{s})$. According to equation (\ref{223}), it can be calculated by
\begin{equation}\label{318}
	\sigma_{R^{A}_{SNe,sel}}=(1-R^{A}_{SNe}(z_{l},z_{s}))\sqrt{(\Delta d_l/d_l)^2+(\Delta d_s/d_s)^2},
\end{equation}
where both $\Delta d_l/d_l$ and $\Delta d_s/d_s$ equal  $5\%$.
$\sigma_{R^{A}_{SGLS}}$  represents the uncertainty of $R^{A}_{SGLS}(z_{l},z_{s})$ and may be computed as follows:
\begin{equation}\label{319}
	\sigma_{R^{A}_{SGLS}}=R^{A}_2 \sqrt{4(\delta \sigma_0)^2+(\delta \theta_E)^2},
\end{equation}
where $\delta \theta_E$  and $\delta \sigma_0$ represent the fractional uncertainty of $\theta_E$ and $\theta_E$ in the sample. Following the SLACS methodology, we fixed the fractional uncertainty of the Einstein radius for all SGLS at 5\%.

The posterior probability density function (PDF) of the parameters is computed using Bayesian statistical methods and the Markov chain Monte Carlo (MCMC) approach to determine the parameters that best suit the $\chi^2$ and the confidence regions of 1$\sigma$ and 2$\sigma$. The probability function is $L\sim exp(-\chi^2/2)$. The settings are optimized using the Python program emcee \citep{for13}. The prior of these parameters is adopted as a uniform distribution: $P(\alpha) = U [-0.2, 0.2]$, $P(\beta) = U [2, 6]$, $P(f_e)=U [0.5, 1.5]$, $P(\varepsilon) = U [-1,1]$, $P(\gamma) = U [0,0.3]$, $P(M_{step}) = U [5,15]$, $P(\tau) = U [0.001,0.3]$. The prior interval of these parameters covers the parameters with 1$\sigma$ interval, which are constrained by \cite{sco18} with a flat  $\Lambda$CDM model.

\subsection{results}
\begin{figure*}
	
	\centering
	\includegraphics[scale=0.4,angle=0]{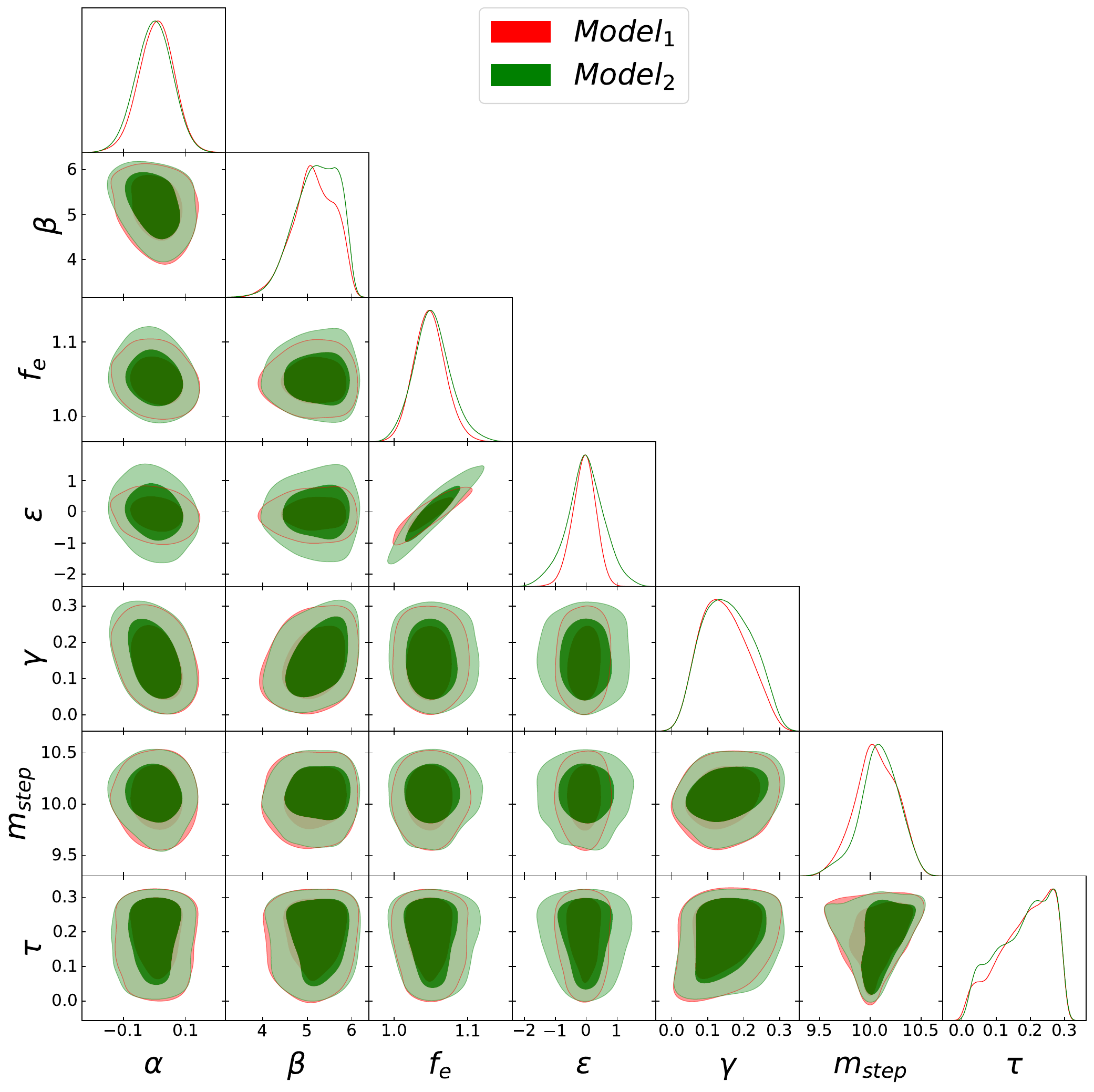}\\
	\caption{The 2-D regions and 1-D marginalized distributions with 1$\sigma$ and 2$\sigma$ contours for the parameters  $\alpha$, $\beta$, $f_e$, $\varepsilon$, $\gamma$, $m_{step}$, and $\tau$ for parametric model of $\delta_M=\varepsilon z$ (the red regions and the red curve) and $\delta_M=\varepsilon \log(1+z)$ (the green regions and the green curve).}
	\label{fig02}
\end{figure*}

We utilize two different kinds of parameterizations to build the triangle figure for the parameters for the $\chi^2$ and the confidence regions of 1$\sigma$ and 2$\sigma$ in figure(\ref{fig02}) for the data that will be used in the study.  Table(\ref{table01}) displays the parameters together with their respective 1$\sigma$ confidence regions. We get to the conclusion that the optimum values of the parameters in $\varepsilon$, as well as their uncertainty in 1$\sigma$, are as follows:
\begin{equation}\label{320}
	\varepsilon=-0.036^{+0.357}_{-0.339},
\end{equation}
for the parametric model $\delta_M=\varepsilon z$, and
\begin{equation}\label{321}
	\varepsilon=-0.014^{+0.588}_{-0.630},
\end{equation}
for the parametric model $\delta_M=\varepsilon\log(1+z)$.

In the results of these two fits, we find that the best-fitted values and their 1$\sigma$ confidence regions of nuisance parameters $\alpha$, $\beta$, and $\gamma$  differ significantly from the previous results \citep{bet14,sco18}, mainly since on the one hand, a model with SGLS was used instead of the standard cosmological model, and on the other hand, the sample of SNe Ia we used is a subsample of the previous published sample, which contains only 240 SNe Ia. However,  the best-fitted values and their 1$\sigma$ confidence regions of nuisance parameters $m_{step}$, and $\tau$ are very close to  the previous work \citep{sco18}. The best-fitted values and its 1$\sigma$ confidence regions the phenomenological free parameter $f_e$ is also very close to  the previous work \citep*{zhou21}. Thus, the parameters $m_{step}$,  $\tau$, and $f_e$ are in line with expectations.

We have found no statistically significant evidence ($\varepsilon=0$ is valid in 1$\sigma$ confidence regions) for the evolution of the absolute magnitudes of SNe Ia. The best value of $\varepsilon$ is negative, but far from a slope of $\sim|0.1|$. This suggests that the hypothesis that absolute magnitudes of SNe Ia do not evolve with redshift is still valid \citep{nor08}. The two parameterized models for $\varepsilon$, equations (\ref{214}) and (\ref{215}), show very different 1$\sigma$ confidence regions, see figure(\ref{fig02}), where the simple linear model gives more centralized 1$\sigma$ confidence regions and tighter constraints on the $\varepsilon$, but, as the analysis in \ref{II.A.} suggests, the second parameterized model is better suited to data containing high redshifts. We can compare the advantages and disadvantages of the two parameterized models by using Akaike weights, and BIC selection weights \citep{anbu04, fan23}, which can be uniformly written in the following form \citep{hu23}
\begin{equation}\label{322}
	P(\alpha)=\frac{\exp{(-XIC_{\alpha}/2)}}{\exp{(-XIC_{other}/2)}+\exp{(-XIC_{\alpha}/2)}},
\end{equation}
where XIC represents the information criterion. Here, we use AIC and BIC \citep{huwa22}, they are defined as
\begin{equation}\label{028}
	AIC=\chi^2+2k,
\end{equation}
and
\begin{equation}\label{029}
	BIC=\chi^2+k\ln(n).
\end{equation}
$k$ is the number of parameters for the two models, n is the number of the data points in section(\ref{II.C.}), and $\chi^2$ is the value in equation(\ref{314}) with the parameters for best fitted. In this work, $k$ and $n$ are the same in both cases, so we only need to consider $\chi^2$. Substituting the numerical values, the relative probabilities for equations (\ref{214}) and (\ref{215}) are 48.4\% versus 51.6\%, this indicates that the second parameterized model is better than the first one.


\section{Conclusions} \label{sec:result}
In this paper, we use the Pantheon SNe Ia sample and the latest SGLS sample to explore the possible evolution of the magnitude with a model-independent method. In this method, the cosmological model is replaced by the SGLS geometric model. To acquire the pure evolution law of absolute magnitude, the parameter of evolution, the model of SGLS, and the SNe Ia light curve are fitted simultaneously with the pair of the SGLS sample and the SNe Ia sample. The fitting results show that no evidence of evolution is found at 1$\sigma$ confidence level with these samples. It means that the SNe Ia is still a powerful and credible tool in studying cosmology with past research methodology. But the best value of the parameter of evolution $\varepsilon$  is a small negative value, which might be related to the sample we took. Exploring this issue in the future may require the use of more and larger samples. What's more, we find that different evolutionary parameter models also affect the fitting results, with parameter models with insignificant evolution with redshift outperforming simple linear models, which may mean that even if more data samples are fitted in the future with $M_B$ having an evolution with redshift, this evolution will be extremely insignificant.

\begin{table*}
	\begin{center}	
		\caption{Constraints on the light-curve parameters and the evolutive parameter of the magnitude of SNe Ia at the 1$\sigma$ confidence levels for two parametric models.} \scalebox{1.0}{
			\begin{tabular}{|c|l|l|}
				\hline
				\diagbox[width=7em,trim=l]{parameters}{model}  &             $\delta_M=\varepsilon z$  &$\delta_M=\varepsilon \ln(1+z)$ \\
				\hline
				$\alpha$  &  $0.004^{+0.057}_{-0.053}$ & $-0.007^{+0.054}_{-0.054}$ \\
				\hline
				$\beta$  &        $5.229^{+0.489}_{-0.524}$&        $5.265^{+0.455}_{-0.485}$\\
				\hline
				$f_e$  &        $1.051^{+0.022}_{-0.020}$   &        $1.053^{+0.025}_{-0.025}$    \\
				\hline
				$\varepsilon$ &        $-0.036^{+0.357}_{-0.339}$   &        $-0.014^{+0.588}_{-0.630}$    \\
				\hline
				$\gamma$ &        $0.147^{+0.083}_{-0.067}$ &        $0.147^{+0.080}_{-0.065}$ \\
				
				\hline
				$m_{step}$ &         $10.079^{+0.208}_{-0.200}$&        $10.130^{+0.188}_{-0.201}$       \\
				\hline
				$\tau$ &         $0.201^{+0.071}_{-0.102}$&        $0.206^{+0.066}_{-0.106}$      \\
				\hline
				$\chi^2$ &       117.518&        117.391      \\
				\hline
				$\chi^2/{d.o.f }$ &        117.518/113 &          117.391/113          \\
				\hline
				AIC(P($\alpha$)) &       131.518(48.4\%)&        131.391(51.6\%)      \\
				\hline
				BIC(P($\alpha$)) &       151.030(48.4\%)&        150.903(51.6\%)       \\
				\hline
		\end{tabular}}
		\label{table01}
	\end{center}	
\end{table*}

\begin{acknowledgements}
	We thank the anonymous referee for constructive comments. This work is supported by  Yunnan Youth Basic Research Projects 202001AU070013, Jiangsu Funding Program for Excellent Postdoctoral Talent (20220ZB59) and Project funded by China Postdoctoral Science Foundation (2022M721561), National Natural Science Foundation of China (No. 11863002), Yunnan Academician Workstation of Wang Jingxiu (No. 202005AF150025), and Sino-German Cooperation Project (No. GZ 1284).
\end{acknowledgements}

\label{lastpage}

\end{document}